\documentclass[article, twocolumn,
   aps, pra,
  amsmath,amssymb,
  longbibliography,
  ]{revtex4-1}
\usepackage{graphicx,color}
\usepackage{amsmath}
\usepackage{natbib}
\usepackage{epsfig}
\begin{document}

\title{Quasi-universal behaviour of shear relaxation times in simple fluids}

\author{S. A. Khrapak}\email{Sergey.Khrapak@gmx.de}
\affiliation{Joint Institute for High Temperatures, Russian Academy of Sciences, 125412 Moscow, Russia}
\author{A. G. Khrapak}
\affiliation{Joint Institute for High Temperatures, Russian Academy of Sciences, 125412 Moscow, Russia}

\begin{abstract}
We calculate the shear relaxation times in four important simple monatomic model fluids: Lennard-Jones, Yukawa, soft-sphere and hard-sphere fluids. It is observed that in properly reduced units, the shear relaxation times exhibit quasi-universal behaviour when the density increases from the gas-like low values to the high-density regime near crystallization. They first decrease with density at low densities, reach minima at moderate densities, and then increase toward the freezing point. The reduced relaxation times at the minima and at the fluid-solid phase transition are all comparable for the various systems investigated, despite more than ten orders of magnitude difference in real systems. Important implications of these results are discussed.       
\end{abstract}

\date{\today}

\maketitle

\section{Introduction}

Shear relaxation time is a key parameter in characterizing the rheological properties of materials and providing insights into their flow and deformation behaviour. The understanding of this parameter is crucial in various scientific, industrial, and engineering applications where the flow and deformation of materials play a significant role. In the Maxwell's model, the shear relaxation time (also referred to Maxwell relaxation time in this case) is defined as the ratio of the viscosity coefficient ($\eta$) to the instantaneous shear modulus ($G_{\infty}$)~\cite{MountainJCP1966},
\begin{equation}\label{Maxwell}
\tau_{\rm M}=\frac{\eta}{G_{\infty}}.
\end{equation}
Maxwell relaxation time plays significant role in liquids. For example it determines the configuration lifetime in the Zwanzig's model of relation between the self-diffusion and viscosity coefficients, i.e. Stokes-Einstein relation~\cite{ZwanzigJCP1983,KhrapakPRE10_2021}. It sets the characteristic frequency for wave propagation in liquids. In particular, it is  closely related to the onset of transverse collective modes in dense liquids and affects the transverse sound speed~\cite{GoreePRE2012,TrachenkoRPP2015,YangPRL2017,BrykPRL2018,KhrapakJCP2019,BaggioliPRE2022}. Thus, the behaviour of the Maxwell relaxation time as the density increases from the ideal gas regime to the freezing point is of great importance and interest.

Actual relaxation times can vary greatly from one system to another. For instance, the relaxation times of argon at its boiling point are of the order of $10^{-12}$ s~\cite{MountainJCP1966}. In viscous liquids relaxation times increase considerably and even more in soft matter systems. Consider for instance underdamped dusty (complex) plasmas -- classical systems of highly charged micron-size particles immersed in a conventional weakly ionized plasma. With typical experimentally measured kinematic viscosities $\eta/m\rho \sim 1$ mm$^2$/s~\cite{NosenkoPRL2004,MorfillPRL2004} and transverse sound velocities $c_t\sim  1$ cm/s~\cite{NosenkoPRL2002,NosenkoPRE2003} we get $\tau_{\rm M}\sim 0.01$ s, ten orders of magnitude longer than in argon. Here $\rho$ is the number density, $m$ is the atomic or particle mass, and the transverse sound velocity is related to the instantaneous shear modulus via $G_{\infty} = m\rho c_t^2$. In overdamped colloidal suspensions even longer relaxation times can be expected. The natural question arises whether this tremendous difference is related to differences in interparticle interactions or differences in characteristic length- and time-scales characterizing different systems.

Information about the viscosity and elastic moduli is sufficient to evaluate the Maxwell relaxation time, and there exist some studies where it has been evaluated for simple fluids at various temperatures and densities. Nevertheless, even for simple fluids, the understanding of general tendencies and magnitudes of relaxation times is rather scarce and fragmented. Mountain and Zwanzig calculated the shear relaxation times of the Lennard-Jones (LJ) fluid at a reduced density up to $\rho^*=0.675$ in conventional LJ units. They observed that $\tau_{\rm M}$ decreases as the density increases up to the higher density investigated. Keshavarzi {\it et al.}~\cite{Keshavarzi2004,Bamdad2005} investigated the density dependence of the shear relaxation time for the soft-sphere (SS) model, LJ model and several real liquids. They found a minimum relaxation time at an intermediate density (e.g. at $\rho^*\simeq 0.7$ in the LJ fluid). Hartkamp {\it at al.} reported a detailed investigation of the density dependence of the stress relaxation function of the Weeks-Chandler-Andersen fluid~\cite{HartkampPRE2013}. A minimum at $\rho^*\simeq 0.7$ was again observed. These studies used system-specific LJ-like unit of time $\sqrt{m\sigma/\epsilon}$ to express the relaxation time, where $\sigma$ is the length-scale and $\epsilon$ is the energy scale of the interaction potential.  

Mainly motivated by strongly coupled dusty plasma systems, the behaviour of the shear relaxation time was investigated in screened Coulomb (Yukawa) fluids~\cite{GoreePRE2012,AshwinPRL2015}. The minimum was observed at a temperature about 10 times the melting temperature  and a density of about 0.001 of the melting density. The relaxation time was expressed in terms of the plasma frequency, which prevents direct comparison of its magnitude with those in other non-plasma-related fluids.     

The calculation of the Maxwellian relaxation times from Eq.~(\ref{Maxwell}) for the hard-sphere (HS) fluid can be controversial. This is related to the fact that the instantaneous shear modulus tends to infinity at all densities in the Zwanzig-Mountain derivation~\cite{ZwanzigJCP1965}, as the HS limit is approached~\cite{HartkampPRE2013,HeyesJCP1994}. This would imply that the Maxwell relaxation time is zero in the HS limit~\cite{HartkampPRE2013,HeyesMolPhys1998}, and that viscous
flow of the HS fluid is perfectly inelastic, which is clearly not the case. Actually, the elastic moduli of the HS fluid are finite~\cite{KhrapakPRE09_2019} and the Maxwell relaxation time can be therefore defined.    

The purpose of this paper is to present a systematic evaluation of the Maxwell relaxation times defined by Eq.~(\ref{Maxwell}) in several simple model fluids with various interactions. We consider the LJ, Yukawa and SS fluids and report new calculations based on recent progress in understanding their transport properties and elastic moduli. Additionally, we provide a new calculation of the shear relaxation time in the HS fluid. In all cases calculations cover the entire density range, from dilute gaseous to dense fluid regime, terminating at the freezing density. We chose the universal system-independent microscopic normalization, so that relaxation times in different fluids can be easily compared.          
It is shown that the Maxwell shear relaxation time behaves universally: it decreases at low densities, reaches a minimum at an intermediate density and then increases towards the freezing point. In the dense fluid regime, corresponding to the densities between the minimum of $\tau_{\rm M}$ and the freezing density, the reduced relaxation times are numerically close for the various simple fluids investigated.

\section{Relaxation times}

In the following we calculate the Maxwellian relaxation time from its definition of Eq.~(\ref{Maxwell}) for four popular important models of simple fluids: Lennard-Jones fluid, screened Coulomb (Yukawa) fluid, soft-sphere fluid, and hard-sphere fluid. The details of these calculations are provided below. 

To facilitate comparison between different fluids we use system-independent macroscopically reduced units throughout the paper. The unit of distance is $\Delta=\rho^{-1/3}$, where $\rho$ is the atom number density; the unit of velocity is the thermal velocity $v_{\rm T}=\sqrt{T/m}$, where $T$ is the temperature in energy units ($\equiv k_{\rm B}T$) and $m$ is the atomic mass; hence the unit of time is $\Delta/v_{\rm T}$. The viscosity coefficient and the instantaneous shear modulus are expressed in units of $mv_T/\Delta^2$ and $\rho T$, respectively. Thus, the main reduced units are
\begin{equation}
\tau_{\rm M}^*=\frac{\tau_{\rm M}v_T}{\Delta}, \quad \eta^* = \frac{\eta\Delta^2}{mv_{\rm T}}, \quad G_{\infty}^*=\frac{G_{\infty}}{\rho T}. 
\end{equation}        

\subsection{Lennard-Jones fluid}  

The pairwise LJ interaction potential is 
\begin{equation}
\phi(r)= 4\epsilon\left[\left(\frac{\sigma}{r}\right)^{12}-\left(\frac{\sigma}{r}\right)^6\right], 
\end{equation}
where $\epsilon$ and $\sigma$ are the energy and length scales, as noted above. The conventional LJ reduced units for temperature and density are $T^*=T/\epsilon$ and $\rho^*=\rho\sigma^3$, respectively.  

\begin{figure}
\includegraphics[width=8.5cm]{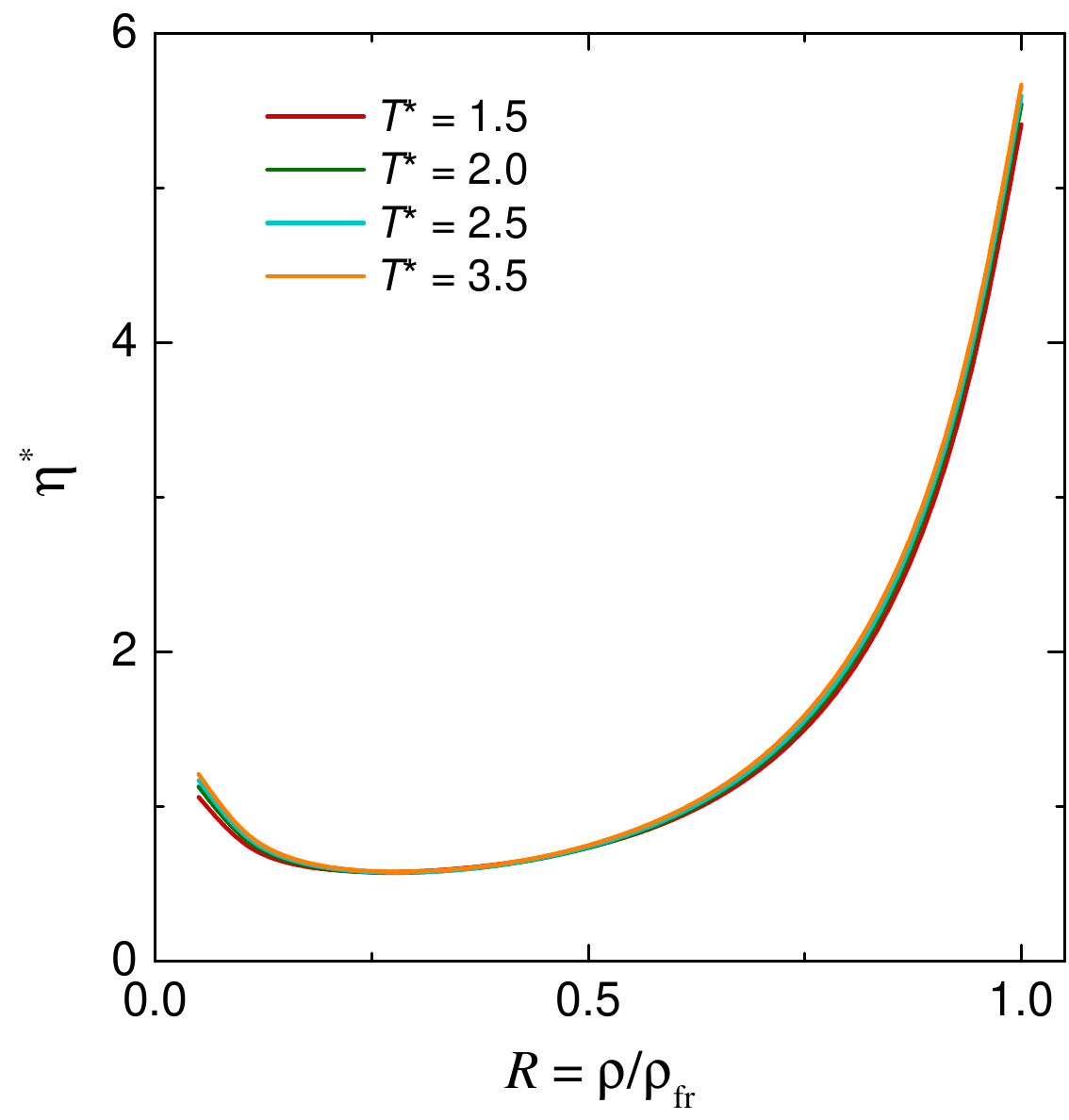}
\caption{(Color online) Reduced shear viscosity coefficient $\eta^*$ of the Lennard-Jones fluid as a function of the FDS parameter ${\mathcal R}=\rho/\rho_{\rm fr}$ along several isotherms ($T^*=1.5$, $2.0$, $2.5$, $3.5$), calculated using the modified excess entropy scaling approach of Ref.~\cite{BellJPCB2019}. Note excellent freezing density scaling of $\eta^*$. }
\label{Fig1}
\end{figure}
 
We evaluate the reduced shear viscosity coefficient $\eta^*$ along several LJ isomorphs using the modified excess entropy scaling approach presented in Ref.~\cite{BellJPCB2019}. Recently discovered freezing density scaling (FDS) of transport coefficients in LJ and related systems~\cite{KhrapakPRE04_2021,KhrapakJPCL2022,KhrapakJCP2022_1,HeyesJCP2023,KhrapakJCP2024} implies that $\eta^*$ should be a quasi-universal function of the density divided by its value at the freezing point, $\rho/\rho_{\rm fr}$, where $\rho_{\rm fr}(T)$ is the freezing density (a somewhat different density scaling relation was reported earlier in Ref.~\cite{SaijaJCP2001}). Figure~\ref{Fig1} provides excellent illustration of the validity and quality of the FDS in the LJ fluid.

\begin{figure}
\includegraphics[width=8.5cm]{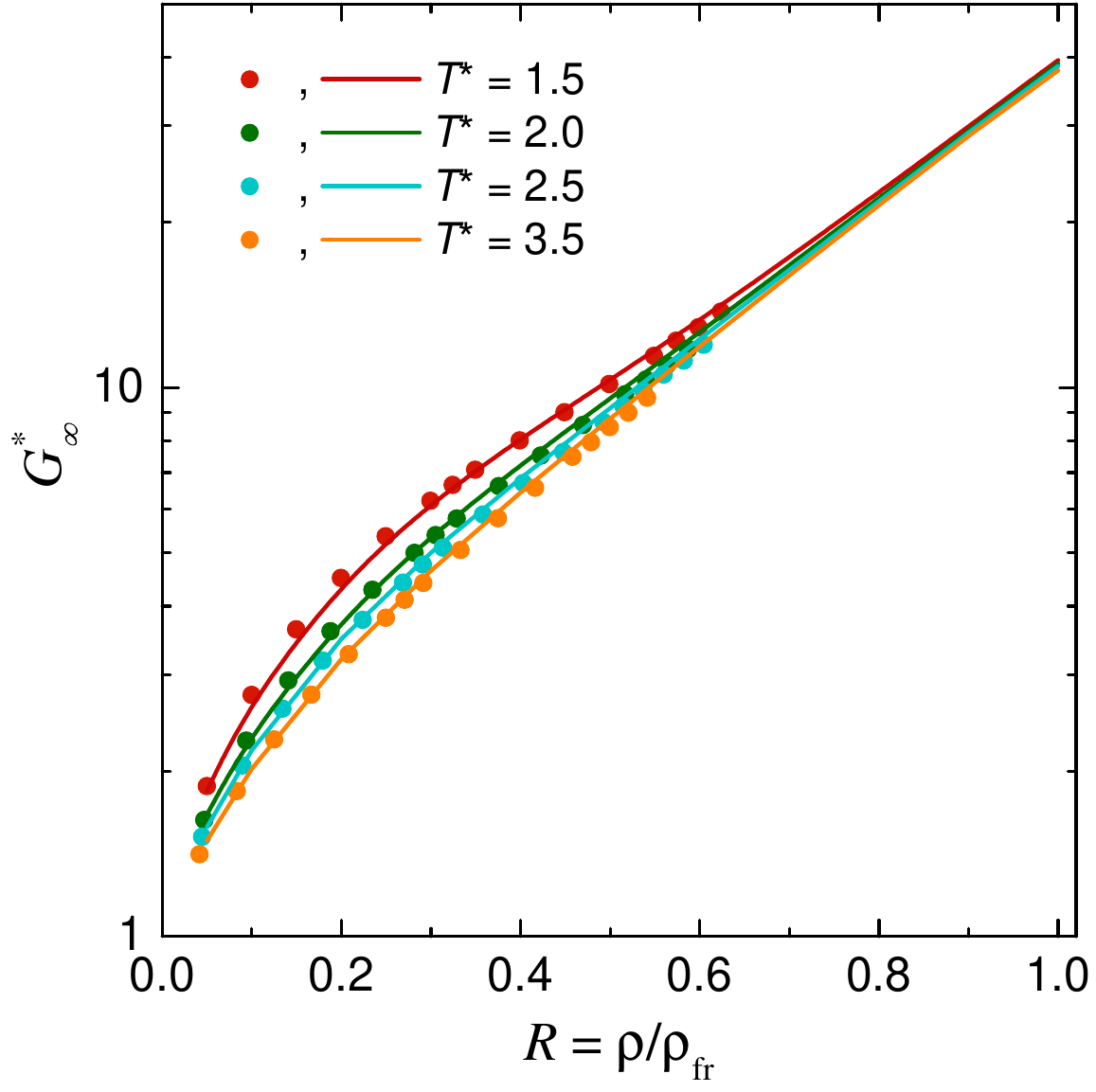}
\caption{(Color online) Reduced instantaneous shear modulus $G_{\infty}^*$ of the LJ fluid as a function of the FDS parameter ${\mathcal R}=\rho/\rho_{\rm fr}$ along several isotherms ($T^*=1.5$, $2.0$, $2.5$, $3.5$). The dotes denote the original calculation by Zwanzig and Mountain~\cite{ZwanzigJCP1965}. The solid lines are our own calculations using the excess energy and pressure from Thol {\it et al}. EoS~\cite{Thol2016}.}
\label{Fig2}
\end{figure}

The instantaneous shear modulus is evaluated using the Zwanzig-Mountain approach~\cite{ZwanzigJCP1965}
\begin{equation}\label{Ginf}
G_{\infty}^*=1+\frac{2\pi \rho}{15 T}\int_0^{\infty}dr \; r^3 g(r)\left[r \phi''(r)+4\phi'(r)\right], \
\end{equation}
where $g(r)$ is the radial distribution function (RDF). The first term is the kinetic contribution. It dominates at low densities corresponding to the dilute gas regime. The second term is the potential (excess) contribution, which dominates in the dense fluid regime. There are two ways of calculating $G_{\infty}^*$. One can perform integration directly, but this requires knowledge of $g(r)$ at each state point investigated. Alternatively, one can use the relation of the integral in Eq.~(\ref{Ginf}) to the excess energy and 
pressure~\cite{ZwanzigJCP1965,KhrapakMolecules2020,KhrapakPRE12_2023} of the LJ system. We use the second option combined with the equation of state developed by Thol {\it et al}.~\cite{Thol2016}. The results are shown in Fig.~\ref{Fig2} by solid curves. The solid circles correspond to the original calculation by Zwanzig and Mountain~\cite{ZwanzigJCP1965}. The agreement is rather good. The freezing density scaling of the reduced instantaneous shear modulus is not so impressive as compared to the viscosity coefficient in Fig.~\ref{Fig1}. Still, at high densities, ${\mathcal R}\gtrsim 0.6$, the convergence is remarkable.

\begin{figure}
\includegraphics[width=8.5cm]{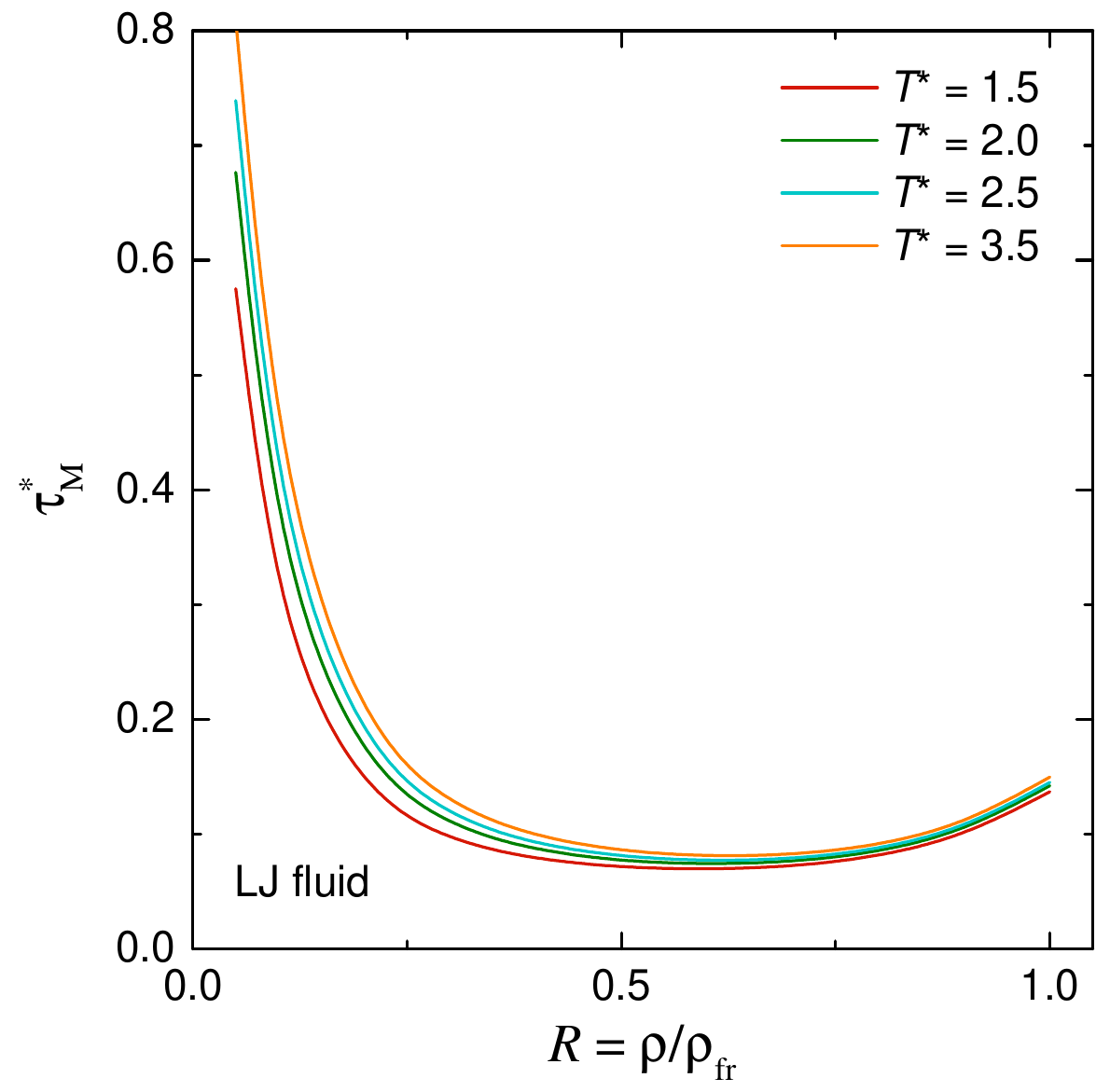}
\caption{(Color online) Reduced shear relaxation time of the LJ fluid as a function of the FDS parameter ${\mathcal R}=\rho/\rho_{\rm fr}$ along several isotherms ($T^*=1.5$, $2.0$, $2.5$, $3.5$).}
\label{Fig3}
\end{figure}

The calculated shear relaxation time as a function of the FDS parameter ${\mathcal R}=\rho/\rho_{\rm fr}$ along different isotherms is plotted in Fig.~\ref{Fig3}. It is observed that $\tau_{\rm M}^*$ is not completely universal in this figure, and the curves for different $T^*$ do not fully overlap. Nevertheless, the convergence improves as the density increases and becomes remarkable as the minimum in $\tau_{\rm M}^*$ is reached. 

The quasi-universal freezing density scaling of the shear viscosity, instantaneous shear modulus, and relaxation time in LJ fluid is not unrelated to the excess entropy scaling, hidden scale invariance, and isomorph theory~\cite{DyreJPCB2014}. It appears that the lines of constant FDS parameter are characterized by approximately constant values of excess entropy~\cite{KhrapakJPCL2022,KhrapakJCP2022_1}. Our results illustrate a special case of a more general observation from Ref.~\cite{KnudsenPRE2021}, where it is shown that the shear viscosity, the rigidity shear modulus and the Maxwell relaxation time in the LJ fluid are all isomorph invariants, including their wave-number dependence.

\subsection{Screened Coulomb (Yukawa) fluid}

The pairwise screened Coulomb repulsive interaction  potential  (also referred to as Debye-H\"uckel or Yukawa potential) is
\begin{equation}\label{Yukawa}
\phi(r)=\frac{\epsilon\sigma}{r}\exp\left(-\frac{r}{\sigma}\right),
\end{equation}
where $\epsilon$ and $\sigma$ are again the energy and length scales. Yukawa potential is often employed as a reasonable first approximation for actual  interactions between charged particles in a polarizable medium, such as complex (dusty) plasmas and colloidal suspensions~\cite{FortovUFN,FortovPR,
IvlevBook,ChaudhuriSM2011,KonopkaPRL1997,KonopkaPRL2000}. In this case $\sigma$ is the so-called screening length, which in the simplest situation is equal to the Debye radius of the screening medium (plasma). The product $\epsilon \sigma$ is equal to the product of electrical charges of the two interacting particles, $\epsilon \sigma=q^2$, if their charges are equal.  Yukawa systems are then described by the two dimensionless parameters: the Coulomb coupling parameter $\Gamma=q^2/aT$ and the screening parameter $\kappa=a/\sigma$, where $a=(4\pi\rho/3)^{-1/3}$ is the Wigner-Seitz radius. The Yukawa potential is purely repulsive and, hence, there are no gas-liquid phase transition, gas-liquid coexistence, critical and gas-liquid-solid triple points. There is still a fluid-solid phase transition at sufficiently strong coupling. The location of the freezing line $\Gamma_{\rm fr}(\kappa)$ for $\kappa<5$ is tabulated in Ref.~\cite{HamaguchiPRE1997} and simple practical fits are available~\cite{VaulinaPRE2002}.    

\begin{figure}
\includegraphics[width=8.5cm]{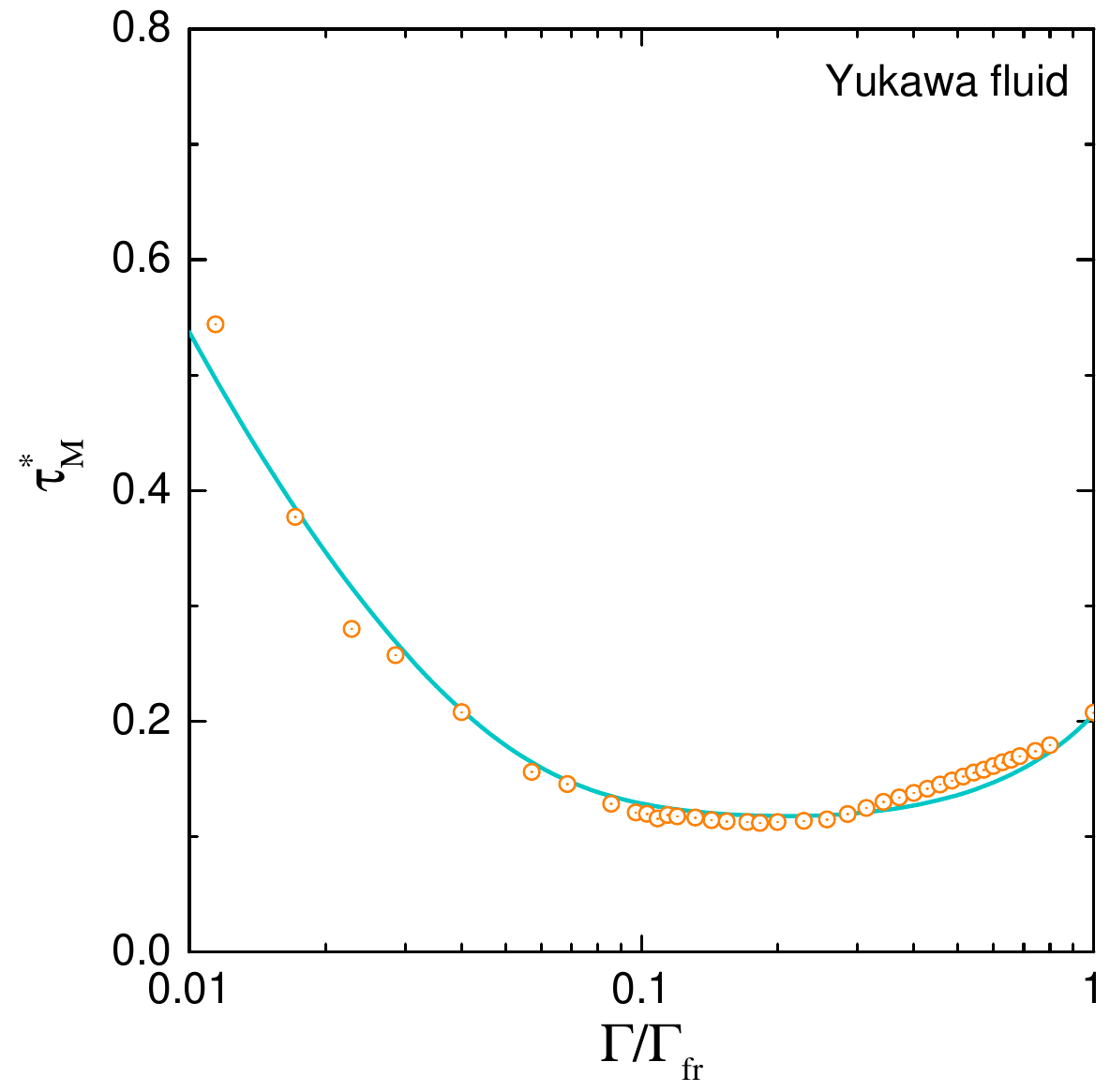}
\caption{(Color online) Reduced shear relaxation time of the Yukawa fluid as a function of the reduced coupling parameter $\Gamma/\Gamma_{\rm fr}$. The symbols correspond to the one-component plasma limit ($\kappa=0$).}
\label{Fig4}
\end{figure}

Transport properties of Yukawa fluids are relatively well known. Accurate data for the shear viscosity coefficients are tabulated for instance in Refs.~\cite{DonkoPRE2008,DaligaultPRE2014}. Here we use a simple practical formula from Ref.~\cite{KhrapakAIPAdv2018}, which allows estimation of the shear viscosity coefficient in a rather extended parameter regime. Moreover, the reduced shear viscosity coefficient appears approximately quasi-universal function of the reduced coupling parameter $\Gamma/\Gamma_{\rm fr}$. The instantaneous shear modulus of strongly coupled Yukawa fluids and solids has also attracted considerable attention~\cite{KhrapakPoP10_2019,KhrapakPoP02_2020,KozhberovPoP2022}. Here we use recent results from Ref.~\cite{YuPRE2024}, demonstrating that $G_{\infty}^*$ is also a quasi-universal function of $\Gamma/\Gamma_{\rm fr}$. As a results, the shear relaxation time of Yukawa fluids is itself a quasi-universal function of $\Gamma/\Gamma_{\rm fr}$. The obtained dependence is shown in Fig.~\ref{Fig4} with a solid curve.     

The symbols in Fig.~\ref{Fig4} correspond to the infinite screening length limit -- the one-component plasma model~\cite{BrushJCP1966,BausPR1980}. To evaluate the Maxwell relaxation time we use molecular dynamics results for the shear viscosity coefficient tabulated in Ref.~\cite{DaligaultPRE2014}. The reduced instantaneous shear modulus is evaluated from $G_{\infty}^*\simeq 1+0.12\Gamma$, based on the results from Ref.~\cite{KhrapakPoP10_2019}, which are consistent with those in Ref.~\cite{YuPRE2024}. The coupling parameter at the fluid-solid phase transition in one-component plasma is $\Gamma_{\rm fr}\simeq 175$~\cite{DubinRMP1999,KhrapakCPP2016}.      

\subsection{Soft-sphere fluid}

\begin{figure}
\includegraphics[width=8.5cm]{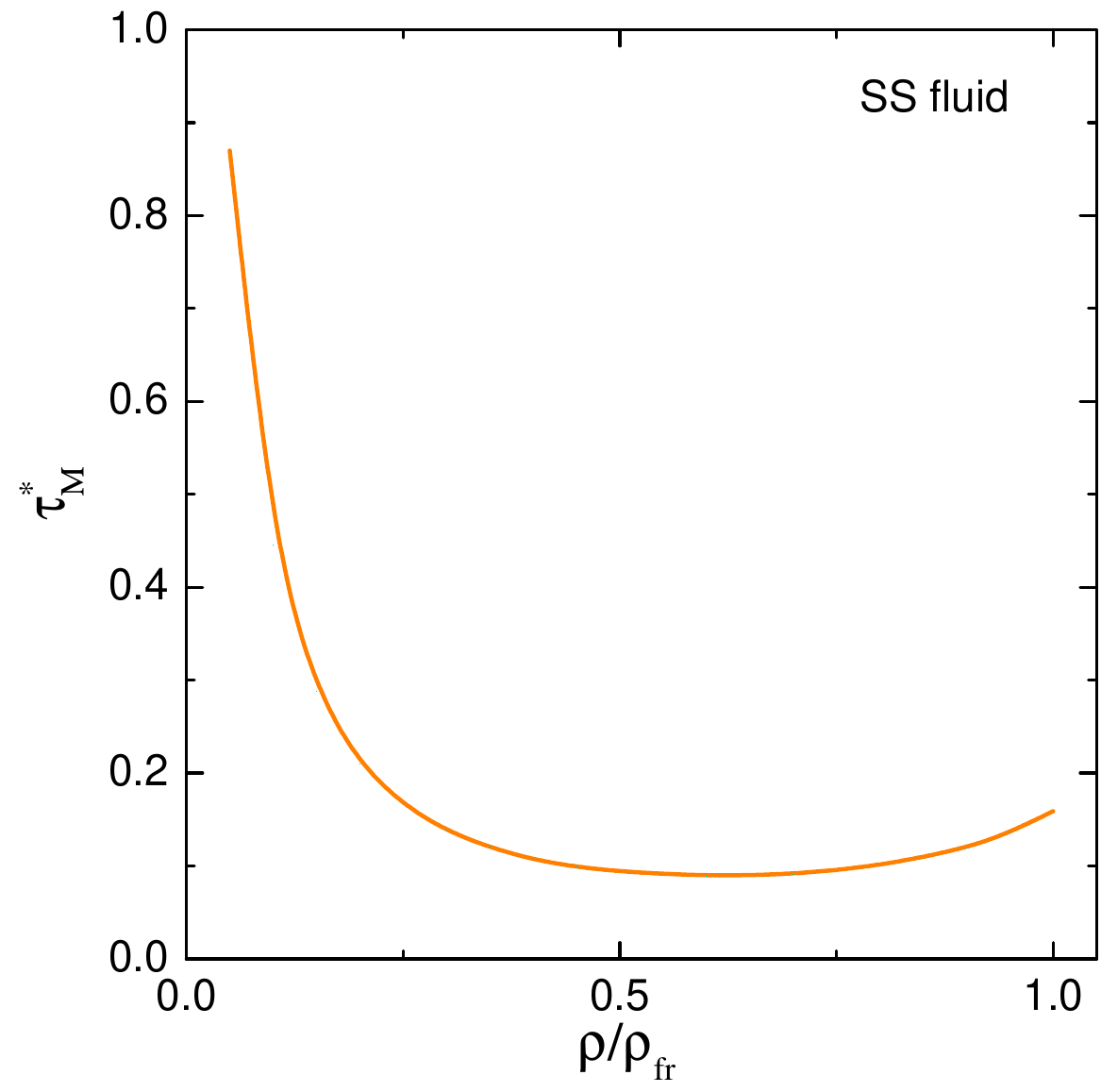}
\caption{(Color online) Reduced shear relaxation time of the soft-sphere fluid as a function of the FDS parameter ${\mathcal R}=\rho/\rho_{\rm fr}$.}
\label{FigSS}
\end{figure}

The soft-sphere interaction potential is defined as
\begin{equation}
\phi(r)=\epsilon\left(\frac{\sigma}{r}\right)^{12},
\end{equation}
where as usually $\epsilon$ and $\sigma$ are the energy and length scales.
This potential is purely repulsive and therefore exhibits only the fluid-solid phase transition at sufficiently high density or low temperature. The properties of the systems are usually described by the single parameter, the effective packing fraction $\gamma=\rho^*(T^*)^{-1/4}$. The thermodynamic and transport properties of the SS model are relatively well known. Our present calculation involves the following steps. The shear viscosity is described by the fitting formula from the Supplementary Information of Ref.~\cite{BellPNAS2019}, based on numerical results from Ref.~\cite{FominJETPLett2012}. The reduced instantaneous shear modulus can be easily related to the compressibility factor $Z=P/\rho T$ in the SS model via~\cite{KhrapakSciRep2017}
\begin{equation}
G^*_{\infty}=1+\frac{9}{5}\left(Z-1\right),
\end{equation}
where $P$ is the pressure. For the compressibility factor we use the 8th order virial expansion from Ref.~\cite{PieprzykPRE2014}. The freezing packing fraction is tabulated in the same work. The resulting dependence of the reduced Maxwell relaxation time on the FDS parameter is shown in Fig.~\ref{FigSS}.

\subsection{Hard-sphere fluid}

The HS interaction potential is extremely hard and short ranged. The interaction energy is infinite for $r<\sigma$ and is zero otherwise, where $\sigma$ is the sphere diameter. The HS system is a very important simple model for the behaviour of condensed matter in its various states~\cite{MuleroBook,ParisiRMP2010,BerthierRMP2011,KlumovPRB2011,DyreJPCM2016}.  

\begin{figure}
\includegraphics[width=8.5cm]{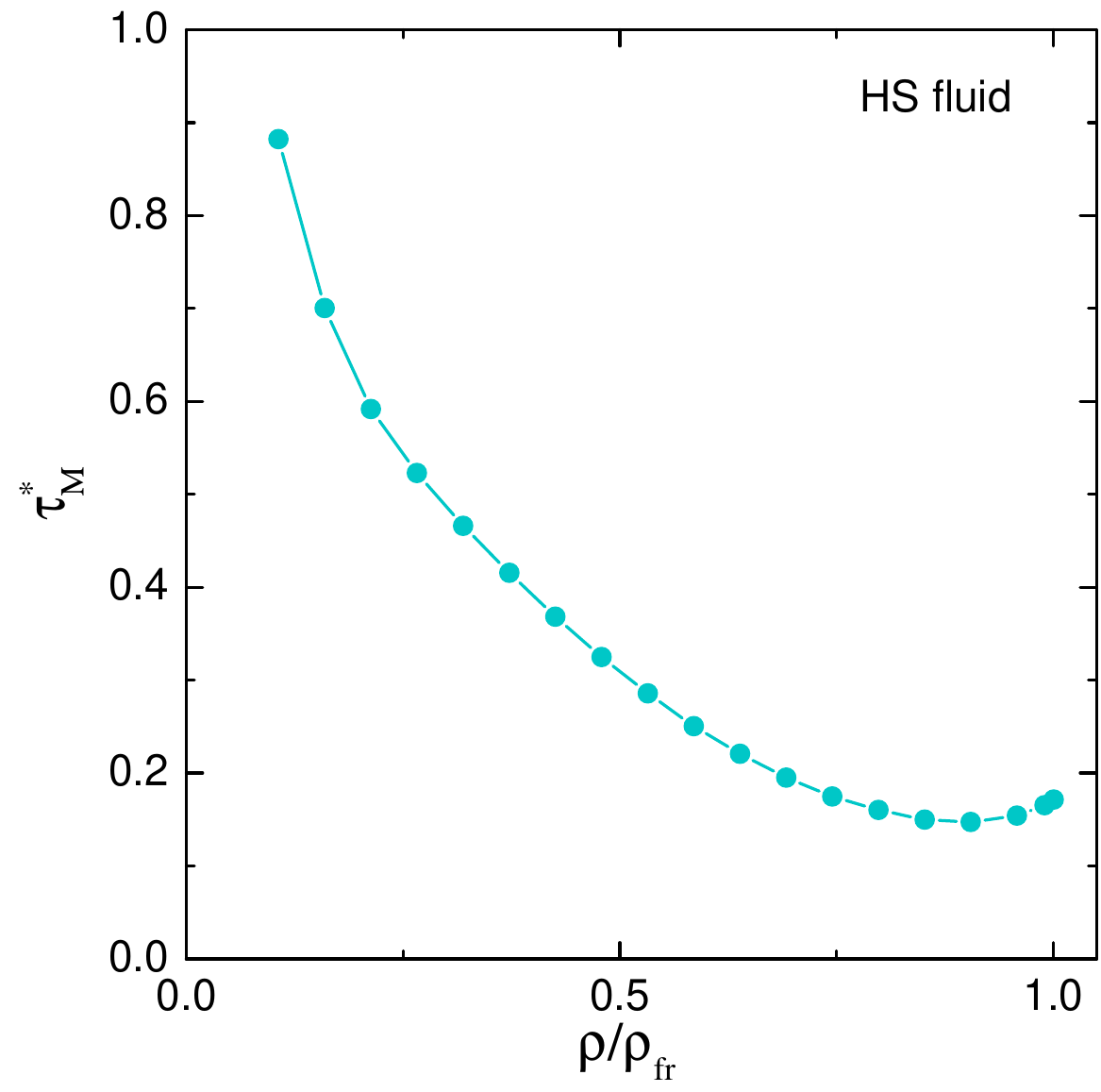}
\caption{(Color online) Reduced shear relaxation time of the hard-sphere fluid as a function of the FDS parameter ${\mathcal R}=\rho/\rho_{\rm fr}$.}
\label{Fig5}
\end{figure}

For the HS system, the thermodynamic and reduced transport properties depend on a single reduced density parameter $\rho^*=\rho\sigma^3$ (the packing fraction, $\pi\rho\sigma^3/6$, is also often used). Transport properties of the HS fluid have been extensively studied (see e.g. Ref.~\cite{MuleroBook} for a review). For our present purpose we use recent MD simulation results for the shear viscosity coefficient reported by Pieprzyk {\it et al}.~\cite{Pieprzyk2019}. The instantaneous shear modulus is somewhat controversial. The point is that the Zwanzig-Mountain expressions such as Eq.~(\ref{Ginf}) imply divergence of the elastic moduli at arbitrary density as the HS limit is approached. This can be illustrated using the inverse power potentials $\phi(r)\propto r^{-n}$ with $n\rightarrow\infty$~\cite{Frisch1966,HeyesJCP1994,GrootMacromolecules1995,HeyesMolPhys1998}.  
However, elastic moduli of the true HS fluid appear finite. This paradox is addressed in a series of recent works~\cite{KhrapakSciRep2017,KhrapakPRE09_2019,KhrapakPRE05_2021}, which conclude that Zwanzig-Mountain approach is not applicable in the limit of very steep HS-like interactions. An alternative used here is based on the Miller's derivation of the elastic moduli for the HS fluid~\cite{MillerJCP1969}, further discussed in Ref.~\cite{KhrapakPRE09_2019}. The derivative of the radial distribution function $g(r)$ at contact (at $r=\sigma$), required in this approach, is estimated using the approximation of Ref.~\cite{TaoPRA1992}. The Maxwell relaxation time calculated in this way is plotted in Fig.~\ref{Fig5} as a function of the FDS parameter.
 
\subsection{Summary}

Now we can summarize our main observations related to Figs.~\ref{Fig3} -- \ref{Fig5}. All systems show the same qualitative trend. In the dilute gaseous regime the relaxation time first decreases as the density increases (note that in the Yukawa system $\Gamma\propto \rho^{1/3}$). Then it reaches a minimum at an intermediate density and increases with density toward the freezing point. The relaxation times at the minima are comparable: $\tau_{\rm M}^*\simeq 0.07-0.08$ for the LJ fluid, $\tau_{\rm M}^*\simeq 0.12$ for the Yukawa fluid, $\tau_{\rm M}^*\simeq 0.9$ for the SS fluid, and $\tau_{\rm M}^*\simeq 0.15$ for the HS fluid.  They are also comparable at the respective freezing points: $\tau_{\rm M}^*\simeq 0.14$ for the LJ fluid, $\tau_{\rm M}^*\simeq 0.2$ for the Yukawa fluid, $\tau_{\rm M}^*\simeq 0.16$ for the SS fluid, and $\tau_{\rm M}^*\simeq 0.17$ for the HS fluid. The increase from the minimum to the freezing point is most pronounced in the case of the LJ fluid. Even in this case, the increase is within a factor of two.       

The minimum in the Maxwell shear relaxation time can be interpreted as the transition between the two asymptotes, one of which corresponding to the relaxation process in dilute gases and the other to the relaxation process in dense liquids. It is tempting to associate the minimum in $\tau_{\rm M}^*$ with the gas-like to liquid-like dynamical crossover (Frenkel line on the phase diagram), which has received considerable attention in recent years~\cite{Simeoni2010, BrazhkinPRE2012,BrazhkinUFN2012,BrazhkinPRL2013,GorelliSciRep2013,ProctorJPCL2019,BellJCP2020}. 
Many criteria for the transition line between the gas-like and liquid-like dynamics have been proposed in the literature. Their reliability and practical convenience may vary. This is not surprising, because the crossover is considered, and thus there is no ``exact demarcation line'' between the two dynamical regimes. The minimum of the Maxwell relaxation time is also an indication of the crossover, similar to the minima of transport coefficients~\cite{GorelliSciRep2013,KhrapakPoF2022}. It is however shifted to somewhat higher densities in comparison to many other proposed indicators. For example, for the LJ fluid most of existing indicators of the gas-like to liquid-like dynamical crossover (e.g. in terms of the excess entropy, minima in reduced shear viscosity and thermal conductivity coefficients, minima of the kinematic viscosity, intersection of asymptotes in the Stokes-Einstein relation) locate it at a FDS parameter ${\mathcal R}=\rho/\rho_{\rm fr}\sim 0.3$. Maxwell relaxation time attains a minimum at a reduced density about two times higher, $\rho/\rho_{\rm fr}\sim 0.6$ (see Fig.~\ref{Fig3}). Similarly, for the Yukawa fluid the gas-like to liquid-like dynamical crossover has been recently located at $\Gamma/\Gamma_{\rm fr}\simeq 0.05$~\cite{YuPRE2024,HuangPRR2023}, while according to Fig.~\ref{Fig4} the minimum in the Maxwell relaxation time is reached at $\Gamma/\Gamma_{\rm fr}\sim 0.2$. For the HS fluids the minimum corresponds to the density rather close to the fluid-solid phase transition density, while the minima in viscosity and thermal conductivity coefficients correspond to $\rho/\rho_{\rm fr}\sim 0.3$~\cite{KhrapakPoF2022}. Thus, the minimum in the Maxwell relaxation time seems to be shifted to the dense fluid regime, where the dynamical picture is dominated by a solid-like vibrational motion~\cite{KhrapakMolecules12_2021,KhrapakPhysRep2024}. 

%It is interesting to note that the reduced Maxwell relaxation times at the gas-like to liquid-like dynamical crossover of the LJ and Yukawa fluids are numerically close to those at the fluid-solid phase transition of these fluids. This correlation is, however, violated in the HS fluid, according to the present calculation.}       

\section{Relaxation times at freezing and some implications}

We observe that as the freezing point (or line) is approached, the reduced relaxation times are all comparable, $\tau_{\rm M}^*\simeq 0.18 \pm 0.04$. We summarize results of our calculations for different systems in Table~\ref{Tab1} and discuss some aspects related to this quasi-universality. 

\begin{table}
\caption{\label{Tab1} Calculated reduced properties of the Lennard-Jones, Yukawa, soft-sphere, and hard-sphere melts: shear viscosity ($\eta^*$), instantaneous shear modulus ($G_{\infty}^*$), relaxation time ($\tau_{\rm M}^*$), wave number corresponding to the onset of transverse collective modes ($k_{\rm gap}^*$), self-diffusion coefficient ($D^*$), ratio of diffusion and Maxwell relaxation times ($\tau_{\rm D}/\tau_{\rm M}$), product of the Einstein frequency and Maxwell relaxation time ($\Omega_{\rm E}\tau_{\rm M}$). }
\begin{ruledtabular}
\begin{tabular}{lccccccc}
System & $\eta^*$ & $G_{\infty}^*$ & $\tau_{\rm M}^*$ & $k_{\rm gap}^*$ & $D^*$ & $\tau_{\rm D}/\tau_{\rm M}$ & $\Omega_{\rm E}\tau_{\rm M}$     \\ \hline
Lennard-Jones &  $5.5$  & $40$ & $ 0.14$ & $ 0.56$ & $ 0.027$ & $ 44$ & $ 2.38$ \\  
Yukawa & $4.8$ & $23$ & $0.21$ & $0.50$ & $0.029$ & $27$ & $3.57$ \\
soft-sphere & $5.7$ & $36$ &  $0.16$ & $0.53$ & $0.027$ & $40$ & $2.75$ \\
hard-spheres & $6.8$ & $40$ & $0.17$ & $0.47$ & $0.025$ & $39$ & $2.89$ \\ 
\end{tabular}
\end{ruledtabular}
\end{table}   

The first three columns with numerical data present the reduced viscosity coefficient $\eta^*$, instantaneous shear modulus $G_{\infty}^*$, and the Maxwell relaxation time $\tau_{\rm M}^*$, as calculated in this work. Note that for the HS system, the fluid-solid phase transition corresponds to the freezing point, while for the LJ and Yukawa systems this is a line on the phase diagram. In the latter case we present average values for the range investigated. The variation are present, but are not very pronounced. For example, in the LJ case, $\tau_{\rm M}^*$ increases from $0.14$ to $0.15$ as the reduced temperature increases from $T^*=1.5$ to $T^*=3.5$.  

The fourth column gives the values of the reduced wave-numbers corresponding to the onset of the transverse collective mode in the considered melts. Dense fluids do support the transverse (shear) waves, but only at sufficiently short length scales, so that the so called $k$-gap -- zero frequency non-propagating domain at low values of the wave-number $k$ is present~\cite{TrachenkoRPP2015,YangPRL2017,BolmatovPCL2015,GoreePRE2012,MurilloPRL2000,
OhtaPRL2000,BrykJCP2017,KhrapakIEEE2018,KhrapakJCP2019,BrykPRL2018}. Within the framework of generalized hydrodynamics, the long-wavelength dispersion relation can be written as~\cite{TrachenkoRPP2015,OhtaPRL2000,KawPoP2001}
\begin{equation}\label{transdisp}
\omega\simeq\sqrt{c_t^2k^2-\frac{1}{4\tau_{\rm M}^2}},
\end{equation}     
where $c_t$ is the transverse sound velocity.
This may not be the best approach to describe the actual dispersion of transverse waves~\cite{KryuchkovSciRep2019}, but it provides a simple way to approximately estimate their cutoff value $k_{\rm gap}$. From Eq.~(\ref{transdisp}) it is obvious that
\begin{equation}
k_{\rm gap}^*\simeq \frac{1}{2c_t^*\tau_{\rm M}^*}.
\end{equation}  
Note that in reduced units $G_{\infty}^*=(c_t^*)^2$, and this allows direct evaluation of $k^*_{\rm gap}=k_{\rm gap}\Delta$. The results presented in Tab.~\ref{Tab1} demonstrate quasi-universal values of the cutoff values,  $k_{\rm gap}^*\simeq 0.50\pm 0.06$ for the systems considered.   

Next column presents the reduced diffusion coefficients, $D^*=D/(v_{\rm T}\Delta)$, of considered melts. These are calculated from the reduced viscosity coefficients using the Stokes-Einstein relation without the hydrodynamic diameter~\cite{CostigliolaJCP2019}, which reads $\eta^*D^*=\alpha_{\rm SE}$ in reduced units. Following Ref.~\cite{KhrapakPRE10_2021} we take $\alpha_{\rm SE}=0.14$ for the Yukawa fluid, $\alpha_{\rm SE}=0.15$ for the LJ fluid, $\alpha_{\rm SE}=0.17$ for the HS fluid. For the SS fluid we take the value $\alpha_{\rm SE}=0.15$~\cite{HeyesPCCP2007}.

We can estimate the characteristic structure relaxation time defined as a characteristic time needed for an atom to diffuse over an average interatomic separation. This relaxation time can be called the diffusion time and defined from the condition $\Delta^2\simeq 6 D\tau_{\rm D}$. In reduced units this results in $\tau_{\rm D}^*\simeq 1/6D^*$. Numerical values for the ratios $\tau_{\rm D}/\tau_{\rm M}$ of the considered melts are given in the sixth column of Tab.~\ref{Tab1}. The presented results indicate that there is a huge separation between the structure relaxation and individual atom dynamical relaxation time scales, justifying the main assumption behind the vibrational model of transport processes in dense fluids~\cite{KhrapakMolecules12_2021,KhrapakPhysRep2024}.  

Puosi and Leporini demonstrated that the structural relaxation time $\tau_{\alpha}$ of a glass forming polymer model system is well correlated with the finite
frequency plateau (or ``relaxed'') shear modulus, but not with the instantaneous shear modulus~\cite{PuosiJCP2012}. Since the infinite frequency modulus is much higher than the plateau modulus (see e.g. Fig. 2 from Ref.~\cite{PuosiJCP2012}), the Maxwell relaxation time $\tau_{\rm M}$ is much shorter than $\tau_{\alpha}$. This makes sense, because $\alpha$-relaxation involves several successive particle rearrangements, most probably by crossing over the large energy barriers between different metabasins~\cite{PuosiJCP2012}. On these relatively long time scales, ``relaxed'' elastic response is more relevant than the instantaneous one.  Quite generally, experimentally determined structure relaxation times in glasses are much longer than Maxwell times~\cite{Lancelotti2021}.

Finally, let us estimate the relationship between the characteristic frequency of solid-like oscillations around the temporary equilibrium positions of atoms in melts and the Maxwell relaxation time. The characteristic oscillation frequency can be invoked from the harmonic Einstein model combined with the Lindemann melting criterion~\cite{Lindemann}. Namely, from energy equipartition, the amplitude of harmonic oscillations of an atom can be expressed using the Einstein frequency
\begin{equation}
\frac{1}{2}m\Omega_{\rm E}^2\langle \delta r^2\rangle = \frac{3}{2}T.
\end{equation}  
The amplitude of high-frequency solid-like oscillations of atoms around their equilibrium positions in melts should not be very different from that in a solid at melting. According to the Lindemann melting criterion the later is $\langle \delta r^2\rangle/\Delta^2\simeq 0.01$. Adopting this value we estimate the product $\Omega_{\rm E}\tau_{\rm M}$ and summarize it in the last column of Tab.~\ref{Tab1}. Again, this estimate supports the applicability of the vibrational model of atomic transport in dense liquids~\cite{KhrapakMolecules12_2021,KhrapakPhysRep2024}. The following hierarchy of characteristic times emerges at near-freezing conditions:
\begin{equation}
1/\Omega_{\rm E} < \tau_{\rm M} \ll \tau_{\rm D}.
\end{equation}

\section{Conclusion}

We report calculations of the Maxwell shear relaxation times for four selected simple model fluids: Lennard-Jones, Yukawa, soft-sphere, and hard-sphere fluids. Expressed in reduced units, the relaxation times behave quasi-universally. They decrease with density in dilute gas regime, reach minima at intermediate densities and further increase towards the fluid-solid phase transition. Numerical values of the reduced relaxation times at their minima and at the fluid-solid phase transition are relatively close ($\pm 30\%$ relative deviations despite more than ten orders of magnitude difference in the absolute values).  Variation of the relaxation time with density in the dense fluid regime is not very large. Maximal ratio of relaxation times at freezing and at the minimum occurs in the Lennard-Jones fluid, but even in this case it does not exceed a factor of two. Numerical values of the reduced relaxation time at freezing conditions, $\tau_{\rm M}^*\sim 0.18\pm 0.04$ allows to estimate important parameters of the simple melts considered.
For the potentials studied no systematic role of the interaction softness or the presence of the long-range attraction has been evidenced. The present results can be of interest for researchers in condensed matter, physics of fluids, physics of plasmas, materials science and beyond.
In future, it would be interesting to verify the robustness of our main conclusions against more complicated potentials, such as anisotropic interactions, potentials with a negative curvature, bounded potentials, with more emphasis on molecular liquids and soft matter.  

%\acknowledgments

%\bibliographystyle{aipnum4-1}

\bibliography{SE_Ref}

\end{document}